\documentclass[11pt,twoside,onecolumn]{article}
\usepackage[]{latexsym}
\usepackage{epsfig}
\usepackage{amsmath,amssymb}
\newcommand{\be}{\begin{eqnarray}}
\newcommand{\ee}{\end{eqnarray}}

\def\theequation{\thesection.\arabic{equation}}
\title{\bf The role of exterior Weyl fluids on compact\\ stellar structures in Randall-Sundrum gravity}
\author{J Ovalle,$^{a}$\thanks{Corresponding author: jovalle@usb.ve}
$\,$ 
F Linares,$^{b}$\thanks{linares.francisco@gmail.com}
$\,$ 
A Pasqua,$^{c}$\thanks{toto.pasqua@gmail.com}
$\,$ 
A Sotomayor$^{d}$\thanks{asotomayor@uantof.cl}
\\
\null
\\
$^a${\em Departamento de F\'{\i}sica, Universidad Sim\'on Bol\'ivar}
\\
{\em Apartado 89000, Caracas 1080A, Venezuela}
\\
\\
$^b${\em Departamento de F\'{\i}sica, Universidad de Guanajuato, Le\'on, M\'exico}
\\
\\
$^c${\em Dipartimento di Fisica, Universit\`a di Trieste}
\\
{\em Via Valerio, 2
34127 Trieste,
Italia}
\\
\\
$^d${\em Departamento de Matem\'aticas, Universidad de Antofagasta, Antofagasta, Chile}
}
\begin{document}
\maketitle
\begin{abstract}
In the context of the Randall-Sundrum braneworld, the minimal geometric deformation approach (MGD) 
is used to generate a new physically acceptable interior solution to Einstein's field equations
for a spherically symmetric compact distribution. This new solution is used to elucidate the role
of exterior Weyl stresses from bulk gravitons on compact stellar distributions. We found strong
evidences showing that the exterior dark radiation  ${\cal U}^+$ always increases both the pressure
and the compactness of stellar structures,  and that the exterior ``dark pressure''
${\cal P}^+$ always reduces them.
\end{abstract}
%
%\keywords{Brane-world, Stars, Black holes}
%\pacs{04.50.+h, 04.70.-s, 04.70.Dy}
%
%
%
%
%
%
\section{Introduction}
\setcounter{equation}{0}
\label{intro}
General Relativity (GR) represents, without any doubt, one of the most important achievement of Physics. The predictions 
made for this theory, like the perihelion shift of Mercury, light deflexion, gravitational red shift, gravitational lens, time delay, etc, 
have given to it the honor which deserves as one of the fundamental theories of Physics (for an excelent review on experimental tests of GR see \cite{cmw} and references therein). 
On the other hand, due to the great technological advance during the past years, today we have increasingly powerful instruments which allow 
to obtain accurate measures associate to the evoution of compact objects, which represent excellent laboratories for the study of gravity in the strong regime (see for instance \cite{sopuerta12}). 
This not only serves to test the theory like never before, but also leaves GR as the only reliable gravitational theory to be used in the analysis of 
phenomena occurred in the strong field regime. Likewise, the ability of observing increasingly distant objects deep in the universe, and 
thus with a great gravitational red shift, leads inevitably to the conclusion that only using GR we can obtain an adequate analysis of these phenomena 
\cite{veneziano}, \cite{ucm}. Furthemore, with the recent results shown by PLANCK \cite{planck}, which improve greatly the previous by WMAP \cite{wmap}, 
we can assure that the cosmological models based in GR enjoys a well-deserved and well-earned prestige.
\par
Despite the above facts, there are some fundamental questions associated to the gravitational interaction which GR cannot answer satisfatory. This can be 
broadly grouped in two fundamental issues, which most likely are closely related: 1) The inability of GR to explain satisfactorily the dark matter \cite{DM} and 
dark energy problem without the need of introducing some kind of unknown matter-energy to reconcile what predicts GR with the observed, namely, 
galactic rotation curves and accelerated expansion of the universe. 2) The impossibility, so far, to reconcile GR with the Standard Model of particle physics, 
or equivalently, the inability to quantize GR. This has strongly motivated the searching of a gravitational theory beyond GR that helps to explain satisfactorily 
part of the problems described above. If the new theory is a consistent quantum theory, this should lead to a generalization of GR at low energy, 
being likely this extension of GR at low energy which could accounts for the dark matter and dark energy problems. If the new theory is not a consistent quantum theory for gravity, 
this should also contain GR in a suitable limit, and somehow show greater tolerance to its quantum description.
\par
Extra-dimensional theories, which are mostly inspired by String/M-theory, are among the theories that lead to modifications to GR. 
One of these extra-dimensional theories is the Braneworld (BW) proposed by Randall and Sundrum (RS) \cite{RS} which has been largely studied and which explains, 
so pretty straightforward, one of the fundamental problems of Physics, i.e. the hierarchy problem (see also the ADD model \cite{ADD} and \cite{AADD}). 
This theory reduces the fundamental scale to the weak scale by considering extra-dimensional effects, thus explaining the weakness of 
gravity relative to the other forces. Because of this, its study and impact on GR is fully justified and is of great importance 
\cite{maartRev2004}, \cite{maartRev2010}, although not yet found experimental evidence to support it \cite{atlas2012}, \cite{cms2013}. (Regarding the dark matter 
problem, a good agreement between the BW theoretical predictions and observations was found in Ref  \cite{laszlo2011}).
\par
Despite the great efforts made in recent years and the great understanding of the theory of RS braneworld, we are still far from fully understanding its 
impact on gravity, mainly in self gravitational systems (See for instance Ref \cite {kanti06} for an study on star and black hole solutions from the perspective of the bulk). Although we have a covariant approach that is useful to study many fundamental aspects of the theory 
\cite{SMS}, there are certain key issues that remain unresolved. One of these problems is the necessity to clarify the existence of black holes solutions in RS. 
In order this theory can be considered a physically consistent theory, it must support, at least, black holes solutions. Today we have some evidences indicating 
the existence of black holes solutions in RS theory ~\cite{FW11, stoj11, LBH13}, but an exact solution remains unknown so far. Indeed, solving the full five-dimensional 
Einstein field equations, something which would be tremendously useful, has proven to be an extremely complicated problem (see, e.g.~\cite{cmazza,darocha},
and references therein). Another aspect worth noting is that, even knowing an exact solution, it would produce different solutions in our observed universe, 
since there are many ways to embed a four-dimensional brane in the five-dimensional bulk. There are indeed many different ways of embedding, e.g. isometric,  
conformal,  imbeddings, rigid,  global, local,  analytic  etc. \cite{maia, laszlo2003, laszlo2008, ruth2000}. While it is true that ${Z}_2$ symmetry is popular when considering this point, 
it is far from being a problem widely studied.
\par
Since the complete solution (bulk
plus brane) remains unknown so far, finding exact solutions, or at least physically acceptable solutions to effective Einstein field equations in the 
brane, might be certainly a good guide to clarify some aspects of the five-dimensional geometry and the way our observed universe is 
embedded in it.  Once a solution is found, we could use Campbell-Magaard theorems \cite{campbell, sss} to extend the brane solution through the bulk, 
locally at least (See Ref ~\cite{maia95, dahia03} for some consequences
of Campbell-Magaard theorems to general relativity). To accomplish this, first of all we need to start with a rigorous study of the effective Einstein field equations in the four-dimensional 
brane. A study that, among other things, clarify the role of five-dimensional effects on the effective four-dimensional 
field equations, and above all, a study to consolidate a general methodology based on a critical basic requirement: regain GR at low energies. Fortunately, 
the RS theory has a free parameter, the tension of the brane, which can be used to control this aspect of extreme importance, and which is a nontrivial problem \cite{
jovalle07}. In fact, this fundamental requirement is the basis of the {\it minimal geometric deformation approach} (MGD) \cite{jovalle2009}, 
which has allowed, among other things, to generate physically acceptable interior solutions for stellar systems \cite{jovalle10}, 
physically acceptable exact interior solutions \cite{jovalle207}, to solve the tidally charged exterior solution found in Ref \cite{dadhich} in terms of the 
ADM mass and to study (micro) black hole solutions \cite{covalle1, covalle2}.  
\par
In the RS BW theory, there are two fields filling our four-dimensional vaccum, namely, the dark radiation ${\cal U}$ and dark pressure ${\cal P}$, which have 
an extradimensional origin, and whose effects on stellar structures is not well understood so far. In this paper the MGD approach is used to generate a 
new physically acceptable interior solution to four-dimensional effective Einstein's field equations for a spherically symmetric compact distribution, which 
is used to elucidate the role of exterior Weyl stresses from bulk gravitons on compact stellar distribution. This paper is organized as follows. 
In Section {\bf 2} the Einstein field equations in the brane for a spherically symmetric and static distribution of density $\rho$ and pressure $p$ is reminded, as well as the MGD approach. 
In Section {\bf 3} a new physically acceptable stellar interior solution to Einstein's fields equations is generated by breaking a well known 
general relativistic perfect fluid solution through the MGD approach. In Section {\bf 4} the consequences of dark pressure  ${\cal P}$ and dark 
radiation ${\cal U}$ on stellar structure will be analized through the matching conditions by using two different exterior solutions. In the last section 
the conclusions are presented. 
\section{Field equations and the minimal geometric deformation \\approach.}
In the context of the braneworld, the five-dimensional gravity produces a modification on Einstein's field equations in our (3+1) observed universe, 
the so-called brane. These modifications can be seen through the energy-momentum tensor, 
which now has new terms carrying five-dimensional consequences onto the brane:
\begin{equation}\label{tot}
T_{\mu\nu}\rightarrow T_{\mu\nu}^{\;\;T}
=T_{\mu\nu}+\frac{6}{\sigma}S_{\mu\nu}+\frac{1}{8\pi}{\cal
E}_{\mu\nu}+\frac{4}{\sigma}{\cal F}_{\mu\nu},
\end{equation}
where $\sigma$ is the brane tension, with  $S_{\mu\nu}$ and
$\cal{E}_{\mu\nu}$ the high-energy and  non-local (from the point of view of a brane observer) corrections respectively, and ${\cal F}_{\mu\nu}$ a term which depends on all stresses in the bulk but the cosmological constant. In this paper, only the cosmological constant will be considered in the bulk, hence ${\cal F}_{\mu\nu}=0$, which implies there will be no exchange of energy between the bulk and the brane, and therefore $\nabla^\nu\,T_{\mu\nu}=0$.
%\begin{equation}
% {\cal F}_{\mu\nu}=0\,\rightarrow\, \nabla^\nu\,T_{\mu\nu}=0
%\end{equation}

Using the line element in Schwarzschild-like coordinates
\begin{equation}
\label{metric}ds^2=e^{\nu(r)} dt^2-e^{\lambda(r)} dr^2-r^2\left( d\theta
^2+\sin {}^2\theta d\phi ^2\right)\, ,
\end{equation}
the effective equations in the brane for a spherically symmetric and static distribution, with Weyl stresses in the interior, can be written as \cite{jovalle2009}
\begin{equation}
\label{usual} e^{-\lambda}=1-\frac{8\pi}{r}\int_0^r r^2\left[\rho
+\frac{1}{\sigma}\left(\frac{\rho^2}{2}+\frac{6}{k^4}\cal{U}\right)\right]dr,
\end{equation}
\begin{equation}
\label{pp}\frac{8\pi}{k^4}\frac{{\cal
P}}{\sigma}=\frac{1}{6}\left(G_1^1-G_2^2\right),
\end{equation}
\begin{equation}
\label{uu}\frac{6}{k^4}\frac{{\cal
U}}{\sigma}=-\frac{3}{\sigma}\left(\frac{\rho^2}{2}+\rho
p\right)+\frac{1}{8\pi}\left(2G_2^2+G_1^1\right)-3p,
\end{equation}
\begin{equation}
\label{con1}p'=-\frac{\nu'}{2}(\rho+p),
\end{equation}
with
\begin{equation}
\label{g11} G_1^1=-\frac 1{r^2}+e^{-\lambda }\left( \frac
1{r^2}+\frac{\nu'}r\right),
\end{equation}
\begin{equation}
\label{g22} G_2^2=\frac{1}{4}e^{-\lambda }\left[ 2\nu''+{\nu'}^2-\lambda'\nu '+2 \frac{\left( \nu'-\lambda'\right)
}r\right],
\end{equation}
where $f'\equiv df/dr$ and $k^2=8{\pi}$. The general relativity
is regained when $\sigma^{-1}\rightarrow 0$ and Eq. (\ref{con1})
becomes a linear combination of Eqs. (\ref{usual})-(\ref{uu}).

Despite the fact that Eqs.(\ref{usual})-(\ref{con1}) represent an indefinite system of
equations in the brane, an open problem for which the solution requires  more information of the bulk geometry and a
better understanding of how our 4D spacetime is embedded in the bulk, it is possible to generate the braneworld 
version of every general relativistic solution through the MGD approach \cite{jovalle2009}. 
In this approach, from the point of view of a brane observer, the five-dimensional gravity produces a geometric deformation in the radial metric component given by
\begin{eqnarray}
\label{expect}  e^{-\lambda}=1-\frac{8\pi}{r}\int_0^r r^2\rho
dr\; +\; \underbrace{Geometric\;\;Deformation}_{=f(\nu,\rho,p)}.\nonumber \\
\end{eqnarray}
When a solution of $4D$ Einstein's field equations is considered as a possible solution to the braneworld system, given by Eqs. (\ref{usual})-(\ref{con1}), the geometric deformation in Eq. (\ref{expect}) is minimal, 
and it is given by
\begin{equation}
\label{fsolutionmin} 
f^{*}(r)=\frac{8\pi
}{\sigma}e^{-I}\int_0^r\frac{e^I}{(\frac{\nu'}{2}+\frac{2}{r})}\left(\rho^2+3\rho p\right)dr+\beta(\sigma)\,e^{-I},
\end{equation}
with
\begin{eqnarray}
\label{I} I\equiv
\int\frac{(\nu''+\frac{{\nu'}^2}{2}+\frac{2\nu'}{r}+\frac{2}{r^2})}{(\frac{\nu'}{2}+\frac{2}{r})}dr,
\end{eqnarray}
and $\beta(\sigma)$ a function of the brane tension $\sigma$ which must be zero in the GR limit. 
In the case of {\it interior} solutions, the condition $\beta(\sigma)=0$ has to be imposed to avoid singular solutions at the center $r=0$. 
As it is shown by Eq. (\ref{fsolutionmin}), the geometric deformation $f^{*}(r)$ satisfies $f^{*}(r)\geq\,0$, hence it always reduces the effective interior mass, 
as it is seen further below in Eqs. (\ref{reglambda}) and (\ref{massfunction}). On the other hand, Eq. (\ref{fsolutionmin}) represents a minimal geometric deformation in the sense that all sources of the geometric 
deformation have been removed except those produced by the density and pressure, which are always present in a stellar 
distribution. However, there is an even minimal deformation in the case of a dust cloud $p=0$, but this will not be considered in the present work.
This geometric deformation $f^{*}(r)$ is the source of the anisotropy induced in the brane, whose explicit form may be found through Eq. (\ref{pp}), leading to
\begin{eqnarray}
\label{ppf3}
\frac{48\pi}{k^4}\frac{{\cal P}}{\sigma}=&&
\underbrace{\bigg(G_1^1-G_2^2\bigg)\mid_{\frac{1}{\sigma}=0}}_{=\,0}+f^{*}\bigg(\frac{1}{r^2}+\frac{\nu'}{r}\bigg)
\nonumber \\ &&
-\frac{1}{4}f^{*}\bigg(2\nu''+{\nu'}^2+2\frac{\nu'}{r}\bigg)-\frac{1}{4}{(f^{*})}'\bigg(\nu'+\frac{2}{r}\bigg).
\nonumber \\
\end{eqnarray}
It is clear that this minimal deformation will produce a minimal anisotropy onto the brane.

\section{A stellar solution.}

The Heintzmann solution \cite{Heint} is a well known spherically symmetric stellar solution for a perfect fluid in general relativity. 
This solution satisfies all the elementary criteria for physical acceptability that a stellar solution must satisfy, 
namely, regular at the origin, pressure and density
defined positive, well defined mass and radius, monotonic decrease
of the density and pressure with increasing radius, dominant energy condition satisfied, subliminal sound speed, etc. Because of the above, the Heintzmann solution  represents a very attractive candidate to be considered in the braneworld context through the minimal geometric deformation approach. Hence 
some interesting aspect of the five-dimensional gravity consequences on stellar structures might be elucidated.
 
Let us start by considering the Heintzmann solution for a perfect fluid in general relativity
\begin{equation}\label{heint00}
e^{\nu}=A\,(1+Cr^{2})^3,
\end{equation}

\begin{equation}\label{heint11}
e^{-\lambda_H}=1 - \frac{3Cr^2}{2}\,\,\frac{1+B(1+4Cr^2)^{-1/2}}{1+Cr^2},
\end{equation}
\begin{equation}
\label{density} \rho(r)=\frac{3 C \left[3B \left(1+3 C
   r^2\right)+\left(3+C
   r^2\right) \left(1+4 C
   r^2\right)^{3/2}\right]}{16
   \pi  \left(1+C r^2\right)^2
   \left(1+4 C r^2\right)^{3/2}}
\end{equation}
and
\begin{equation}
\label{pressure} p(r)=\frac{3\,
    C\, \left[ -B(1 + 7\,  C\,
        r^2) + 3\, \left( 1 - C\,
              r^2 \right) \, {\sqrt{1 + 4\, C\,
                r^2}} \right] }{16\, \pi \, {\left( 1 + C\,
              r^2 \right) }^2\, {\sqrt{1 + 4\, C\, r^2}}},
\end{equation}
where $A$, $B$  and $C$ are constants to be determined by matching
conditions. In general relativity, all these constants have specific values. Indeed, they
may be written in terms of the compactness of the distribution, that is, in terms of $M/R$, 
with $M$ and $R$ the mass and radius of the distribution, which are free parameters satisfying the constraint $M/R<4/9$. However, as it is well known,
in the braneworld scenario the matching conditions are modified, consequently there are five-dimensional effects on these constants which must be considered.

In general relativity the second fundamental form, which leads to
$p(r)\mid_{r=R}\,=0$ at the stellar surface $r=R$, produces 
\begin{equation}
\label{sff}B= \frac{3\, \left( 1 - C\, R^2 \right) \, {\sqrt{1 +
4\, C\, R^2}}}{1 + 7\, C\,
    R^2}.
\end{equation}
We will keep the physical pressure vanishing on the surface, even though this condition may be dropped in the braneworld scenario \cite{deru},
\cite{gergely2007}. 

From the point of view of a brane observer, the geometric deformation $f^{*}(r)$ produced by five-dimensional effects ``breaks'' the perfect fluid solution 
represented by Eqs. (\ref{heint00})-(\ref{pressure}), introducing thus imperfect fluid effects through the braneworld solution for the geometric function $\lambda(r)$, 
which is obtained using Eqs. (\ref{heint00}), (\ref{density}) and (\ref{pressure}) in Eq. (\ref{expect}), leading to
\begin{equation}\label{reglambda}
e^{-\lambda(r)}=1-\frac{2\tilde{m}(r)}{r},
\end{equation}
where the interior mass function $\tilde{m}$ is given by
\begin{equation}
\label{massfunction}
\tilde{m}(r)=m(r)-\frac{r}{2}\,f^{*}(r), 
\end{equation}
with $f^{*}(r)$ the {\it minimal geometric deformation} for the Heintzmann solution, given by Eq. (\ref{fsolutionmin}), whose explicit form is obtained using 
Eqs. (\ref{heint00}), (\ref{density}) and (\ref{pressure}) in Eq. (\ref{fsolutionmin}), hence
%\begin{widetext}
\begin{eqnarray}
\label{gr} f^{*}(r)&=&\frac{1}{\sigma}\frac{-9\,
    C^2}{8\pi{r}(1+Cr^2)(2+5Cr^2)^{11/10}} 
\int_0^r \frac{x^2\sqrt[10]{2+5 C x^2} }{{\left( 1 + 4Cx^2 \right) }^3{\left(1  + Cx^2 \right) }^2}
\nonumber \\ &&
\left[ -9  + 3 (6B^2 -35 )Cx^2 
 + (117\, B^2 -394 )\,C^2x^4  \right.
\nonumber \\ &&
           + 3\,( 63\, B^2 -136 )\,C^3x^6  + 288\, C^4\,
          x^8\, + 128\, C^5\,
          x^{10} 
\nonumber \\ &&
+3B\, {\sqrt{1 + 4Cx^2}}
\left.( -3  - 13Cx^2+25 C^2x^4
+ 123C^3x^6+28C^4x^8)
 \right]dx.
\nonumber \\
\end{eqnarray}
%\end{widetext}
The function $m(r)$ in Eq. (\ref{massfunction}) is the general relativity interior mass function,
given by the standard form
\begin{equation}
\label{regularmass2} m(r)=\int_0^r 4\pi
r^2{\rho}dr=\frac{3Cr^3}{4}\left[\frac{1+B(1+4Cr^2)^{-1/2}}{1+Cr^2}\right],
\end{equation}
hence the total general relativity mass is obtained
\begin{equation}
\label{regtotmass} M\equiv
m(r)\mid_{r=R}\,=\frac{3CR^3}{4}\left[\frac{1+B(1+4CR^2)^{-1/2}}{1+CR^2}\right].
\end{equation}
On the other hand, it can be shown by Eqs. (\ref{sff}) and (\ref{gr}) that the geometric deformation $f^{*}(r)$ 
depends only on the parameter $C$, which has a well defined expression in terms of the compactness in general relativity. 
The general relativity expressions for $A$ and $C$, which from now on will be called $A_0$ and $C_0$ respectively, are found in terms of $M$ and $R$ by 
the continuity of the metric at the stellar surface, where, of course, the Schwarzschild's exterior solution must be used. 
Thus, by considering the temporal component of the metric, we have
\begin{equation}
\label{RegmatchNR1GRxx} A_0(1+C_0R^{2})^{3}=1-\frac{2M}{R}.
\end{equation}
Hence using Eq. (\ref{sff}) in Eq. (\ref{regtotmass}) and Eq. (\ref{RegmatchNR1GRxx}) two simple expression relating $A_0$ and $C_0$ with the compactness are found 
as shown below
\begin{equation}
\label{C0}
 C_0R^2=\frac{M/R}{3-7M/R},
\end{equation}
\begin{equation}
\label{A0}
A_0=\frac{1}{9}\frac{\left(3-7M/R\right)^2}{\left(1-2M/R\right)}, 
\end{equation}
%\begin{equation}
%\label{grex}
% A\,\left(1+C_0\,R^2\right)^2\left(1+7\,C_0\,R^2\right)=1,
%\end{equation}
showing thus that the geometric deformation $f^{*}(r)$ in Eq. (\ref{gr}) can be written in terms of the compactness of the distribution, therefore it may be used as a 
free parameter to model different compact stellar distributions. Indeed, it is found that more compact distributions undergo a higher deformation due to five-dimensional effects. 
In order to obtain the interior Weyl functions ${\cal P}$ and ${\cal U}$, Eqs. (\ref{heint00}), (\ref{density}) and (\ref{pressure}) along with Eq. (\ref{reglambda}) 
and Eq. (\ref{massfunction}) are used in Eq. (\ref{pp}) and Eq. (\ref{uu}), leading to expressions too large to be shown here, but proportional to the geometric deformation.
A numerical analysis carried out inside the stellar distribution, for different densities, shows that the 
anisotropic stress ${\cal P}$ is proportional to the density: the most compact distribution undergoes 
a higher anisotropic effect. This behavior, which has already been observed in the study of uniform stellar distributions \cite{jovalle10},  
can be easily explained in terms of the source of the anisotropy, which is the minimal geometric deformation undergone by the radial metric component, 
explicitly shown through Eq. (\ref{fsolutionmin}). As the Heintzmann solution is a solution to $4D$ Einstein's field equations, it removes all the 
non-local sources from the geometric deformation $f(\nu,\rho,p)$ in the generic expression given by Eq. (\ref{expect}), leaving only the high energy terms shown explicitly in Eq. (\ref{fsolutionmin}), which are quadratic terms in the density and pressure. Hence the higher the density, the more geometric deformation will be produced, and as a consequence the anisotropy induced will be higher for more compact 
distributions. On the other hand, when the scalar Weyl function ${\cal U}$ is analyzed, it is found that this function is more negative for more compact stellar objects. The fact that ${\cal U}$ is always negative means that high energy terms 
always dominate anisotropic terms, which are the two sources for ${\cal U}$, as can be seen through Eq. (\ref{uu}). Next, using two different exterior solutions having a Weyl fluid, the consequences of dark pressure  ${\cal P}$ and dark radiation ${\cal U}$ on stellar structure will be analized through the matching conditions.

\section{The role of exterior Weyl fluids}

It is well known that in general relativity the unique exterior solution for a spherically symmetric distribution is the Schwarzschild metric. This situation 
changes dramatically in the five-dimensional braneworld scenario, where high energy corrections and the presence of the Weyl stresses imply that the exterior solution for a 
spherically symmetric distribution is no longer the Schwarzschild metric. Among all possible exterior solutions known to date, the one obtained by Dadhich, 
Maartens, Papadopoulos and Rezania (DMPR) in Ref. \cite{dadhich}, given by  
\begin{equation}
\label{RegRNmet}
e^{\nu^+}=e^{-\lambda^+}=1-\frac{2\cal{M}}{r}+\frac{q}{r^2},
\end{equation}
\begin{equation}
\label{RegRNmet2} {\cal U}^+=-\frac{{\cal P}^+}{2}=\frac{4}{3}\pi
q\sigma\frac{1}{r^4},
\end{equation}
represents the simplest generalization of the Schwarzschild exterior 
solution in the braneworld. Using our interior solution given by Eqs. (\ref{heint00}) and (\ref{reglambda})
along with the Reissner-N\"{o}rdstrom-like solution Eq. (\ref{RegRNmet}) in the matching conditions at the
stellar surface $r=R$, we have
\begin{equation}
\label{RegmatchNR1}
A(1+CR^{2})^{3}=1-\frac{2\cal{M}}{R}+\frac{q}{R^2},
\end{equation}
\begin{equation}\label{RegmatchNR2}
\frac{2\cal{M}}{R}=\frac{2M}{R}-f^{*}(C)+\frac{q}{R^2}.
\end{equation}
The tidal charge $q$ in Eq. (\ref{RegmatchNR1}) and Eq. (\ref{RegmatchNR2}) is obtained from the second fundamental form, 
which in the braneworld is given by
\begin{equation}
\label{sffxx}
p_R+\frac{1}{\sigma}\left(\frac{\rho_R^2}{2}+\rho_R p_R
+\frac{2}{k^4}{\cal U}_R^-\right)+\frac{4}{k^4}\frac{{\cal
P}_R^-}{\sigma} = \frac{2}{k^4}\frac{{\cal
U}_R^+}{\sigma}+\frac{4}{k^4}\frac{{\cal P}_R^+}{\sigma},
\end{equation}
but in this approach is reduced to
\begin{equation}
 \label{sffmgd}
p_R+\frac{f^*(C)}{8\pi}\left(\frac{\nu'(R)}{R}+\frac{1}{R^2}\right)= \frac{2}{k^4}\frac{{\cal
U}_R^+}{\sigma}+\frac{4}{k^4}\frac{{\cal P}_R^+}{\sigma},
\end{equation}
thus using the DMPR solution given by Eq. (\ref{RegRNmet2}) in Eq. (\ref{sffmgd}) we found
\begin{equation}
\label{sffgen} q/R^4=-\left(\frac{\nu'(R)}{R}+\frac{1}{R^2}\right)f^{*}(C)-8{\pi}p_R.         
\end{equation}
which for our interior solution given by Eq. (\ref{heint00}) leads to
\begin{equation}
\label{qNR} q/R^2=-\frac{1+7CR^2}{1+CR^2}f^{*}(C)\, .       
\end{equation}
The expression $f^{*}(C)\equiv f^{*}(R)$ is the minimal geometric deformation at the stellar surface $r=R$. Since the geometric deformation $f^{*}(C)$ 
is always positive, it can be seen from Eq. (\ref{qNR}) that the tidal Weyl charge $q$ is always negative.
%, in consequence the exterior solution will have only one horizon $r_h$ lying outside the Schwarzschild horizon,
%\begin{equation}
% \label{horizon}
%r_{h}={\cal{M}}+\sqrt{{\cal{M}}^2-q}.
%\end{equation}
The constants $\cal{M}$ and $q$ are given in terms of $C$ through Eqs. (\ref{RegmatchNR2}) and (\ref{qNR}) respectively, and
$C$ may be determined by Eq. (\ref{RegmatchNR1}) if $A$ is kept as a
free parameter, which can be used to find a physically acceptable
model. On the other hand, the general relativistic value of $C$, named $C_0$, can be seen by 
 Eq. (\ref{RegmatchNR1}) evalated at $\sigma^{-1}=0$, which leads to Eq. (\ref{RegmatchNR1GRxx}). Hence comparing Eq. (\ref{RegmatchNR1}) with 
Eq. (\ref{RegmatchNR1GRxx}), it is clear that 
the general relativistic values of $C$ has been modified
by bulk effects as
\begin{equation}
\label{RegAC} C(\sigma)=C_0+\delta(\sigma).
\end{equation}
In order to determine $\delta(\sigma)$, Eq. (\ref{RegAC}) is used in Eq. (\ref{RegmatchNR1}), hence  
\begin{equation}
\label{RegmatchNR1Pert} A[1+(C_0+\delta)R^{2}]^{3}=1-\frac{2{\cal
M}}{R}+\frac{q}{R^2}\, .
\end{equation}
Now using Eqs. (\ref{RegmatchNR2}) and (\ref{qNR}) in Eq. (\ref{RegmatchNR1Pert}), and keeping linear terms in $\sigma^{-1}$, we found
\begin{equation}
\label{dNR}
C(\sigma)=C_0+\underbrace{\frac{f^{*}(C_0)}{3\,A \,R^2\,(1+C_0\,R^2)^2}+{\cal
O}(\sigma^{-1})}_{\delta(\sigma)}\, .
\end{equation}
As expected, the modification of $C$ is proportional to the geometric deformation at the surface.

At this stage, we have all the necessary tools needed to examine the braneworld consequences on the physical variables. 
For instance, to see five-dimensional consequences on the pressure $p$ for different distributions, all 
we have to do is to use Eq. (\ref{dNR}) in Eq. (\ref{pressure}) with different values of the positive parameter $A$ ($A<1$). 
As the compactness of the distribution is increased when $A$ decreases, we may obtain the five-dimensional consequences for different compact distributions. 
Figure 1 shows the behavior of the pressure in both the
general relativity and braneworld case. In the braneworld case, two different exterior solutions are considered. It can be seen that the pressure is increased 
by five-dimensional 
effects when the DMPR solution is used. The same result was found in Ref \cite{jovalle10} for a uniform stellar distribution described by a braneworld version of the Schwarzschild's metric, where 
high energy corrections were considered along with Weyl stresses from bulk gravitons, which represents a different result from other braneworld 
solution \cite{germ}, where only high energy modifications were considered. The latter strongly suggests that, at least for the Schwarzschild's braneworld model, 
unlike the high energy effects, the Weyl stresses increases the pressure, and that its effects dominate over the high energy effects. However, the specific role played by each Weyl function cannot be elucidated yet, unless one of them is turned off in order to see its consequences, but accomplishing this without breaking the interior braneworld solution is very complicated. The only way to see the individual effects of Weyl functions on stellar structure is through an exterior solution with one Weyl function off. Fortunately, there is a vacuum solution that complies with this feature, namely, 
the vacuum braneworld solution found by Casadio, Fabbri and Mazzacurati (CFM) in Ref. \cite{CFMsolution}, given by
\begin{equation}
 \label{cfm00}
e^{\nu^+}=\left[\frac{\eta+\sqrt{1-\frac{2{\cal M}}{r}(1+\eta)}}{1+\eta}\right]^2,
\end{equation}
\begin{equation}
 \label{cfm11}
e^{\lambda^+}=\left[1-\frac{2{\cal M}}{r}(1+\eta)\right]^{-1},
\end{equation}
\begin{equation}
 \label{cfmp}
\frac{16\pi{\cal P}^+}{k^4\sigma}=-\frac{{\cal M}(1+\eta)\eta}{\eta+\sqrt{1-\frac{2{\cal M}}{r}(1+\eta)}}\frac{1}{r^3},
\end{equation}
\begin{equation}
 \label{cfmu}{\cal U}^+=0,
\end{equation}
where $\eta$ is a constant which measures deviation from the Schwarzschild's solution.
Using our interior solution given by Eq. (\ref{heint00}) and Eq. (\ref{reglambda})
with the CFM solution Eqs. (\ref{cfm00})-(\ref{cfm11}) in the matching conditions at the
stellar surface $r=R$, we have
\begin{equation}
\label{matchcfm1}
A(1+CR^{2})^{3}=\left[\frac{\eta+\sqrt{1-\frac{2{\cal M}}{R}(1+\eta)}}{1+\eta}\right]^2,
\end{equation}
\begin{equation}\label{matchcfm2}
\frac{2\cal{M}}{R}(1+\eta)=\frac{2M}{R}-f^{*}(C),
\end{equation}
with the second fundamental form Eq. (\ref{sffmgd}) leading to
\begin{equation}
\label{matchcfm3} 
\frac{2{\cal M}(1+\eta)\eta}{R\left(\eta+\sqrt{1-\frac{2{\cal M}}{R}(1+\eta)}\right)}=-\frac{(1+7CR^2)}{(1+CR^2)}f^{*}(C).    
\end{equation}
The constants ${\cal M}$ and $\eta$ are obtained in terms of $C$ by Eq. (\ref{matchcfm2}) and Eq. (\ref{matchcfm3}), while $C$ is 
determined through Eq. (\ref{matchcfm1}), where $A$ is kept as a free parameter which again will be useful to modeling distributions with different compactness. Using Eq. (\ref{matchcfm2}) in Eq. (\ref{matchcfm1}) we obtain 
\begin{equation}
\label{matchcfm123}
A(1+CR^{2})^{3}=\left[\frac{\eta+\sqrt{1-\frac{2M}{R}+f^{*}(C)}}{1+\eta}\right]^2.
\end{equation}
Keeping linear terms in $\sigma^{-1}$ and $\eta$, Eq. (\ref{matchcfm123}) may be written as
%\begin{widetext}
\begin{eqnarray}
\label{matchcfm1234}
A(1+CR^{2})^{3}&=&1-\frac{2M}{R}+f^{*}(C)
+2\eta\left(1-\frac{2M}{R}\right)\left(\frac{1}{\sqrt{1-\frac{2M}{R}}}-1\right)
\nonumber \\ &&
+\frac{\eta\,f^{*}(C)}{2}\left(\frac{1}{\sqrt{1-\frac{2M}{R}}}-4\right).
\end{eqnarray}
Comparing this last expression with its general relativistic counterpart given in Eq. (\ref{RegmatchNR1GRxx}), it is clear that 
the general relativistic values of $C$ has been modified
by bulk effects by
\begin{equation}
\label{RegAC2} C(\sigma)=C_0+\delta(\sigma,\eta).
\end{equation}
Using Eq. (\ref{RegAC2}) in Eq. (\ref{matchcfm1234}) and keeping linear terms in $\delta$, we obtain
\begin{equation}
\label{RegAC3} \delta(\sigma,\eta)=\frac{f^{*}(C_0)+2\eta(1-\frac{2M}{R})(\frac{1}{\sqrt{1-\frac{2M}{R}}}-1)+\frac{\eta\,f^{*}(C_0)}{2}\left(\frac{1}{\sqrt{1-\frac{2M}{R}}}-4\right)}{3AR^2(1+C_0R^2)^2}.
\end{equation}
%\end{widetext}
In order to find $\eta=\eta(\sigma)$, Eq. (\ref{matchcfm2}) is used in the second fundamental form Eq. (\ref{matchcfm3}), hence
\begin{equation}
 \label{eta5}
\eta(\sigma)=-\frac{(1+7C_0R^2)\sqrt{1-\frac{2M}{R}+f^{*}(C_0)}}{\frac{2M}{R}(1+C_0R^2)+6C_0R^2f^{*}(C_0)}f^{*}(C_0).
\end{equation}
From the last expression and since the geometric deformation $f^{*}>0$, it can be seen that $\eta<0$, hence the exterior solution has a singular horizon at 
$r=\frac{2{\cal M}}{1-\eta}$ \cite{CFMsolution}. Using Eq. (\ref{eta5}) in Eq. (\ref{RegAC3}) 
and keeping linear terms in $\sigma^{-1}$, after some algebraic manipulation and considering the zero order of Eq. (\ref{matchcfm1234}), which is given by Eq. (\ref{RegmatchNR1GRxx}), it is found that under the CFM exterior solution the deviation of $C$ is given by
\begin{equation}
\label{deltacfm}
 \delta(\sigma)=\frac{1}{3AR^2(1+C_0R^2)^2}\left[\frac{\sqrt{A(1+C_0R^2)^3}-1}{\sqrt{A(1+C_0R^2)^3}+1}\right]f^{*}(C_0),
\end{equation}
which can be written as
\begin{equation}
\label{deltacfm2}
 \delta(\sigma)=\frac{1}{3AR^2(1+C_0R^2)^2}\left[\frac{\sqrt{1-\frac{2M}{R}}-1}{\sqrt{1-\frac{2M}{R}}+1}\right]f^{*}(C_0).
\end{equation}
Using the general relativistic upper bound for the compactness limit of the star, $M/R<4/9$,  it is found from Eq. (\ref{deltacfm2}) that $\delta(\sigma)<0$. The fact that the parameter $C$ is reduced by five-dimensional effects has important consequences. As it is shown in figure 1, the decrease in $C$ due to the CFM exterior solution, unlike the case $\delta(\sigma)>0$ given in Eq. (\ref{dNR}) for the DMPR exterior solution, produces a decrese in pressure. Finally, it is easy to see that both expressions given by Eq. (\ref{dNR}) and Eq. (\ref{deltacfm}) can be written in a concise way, as shown below
\begin{equation}
\label{deltacon2}
 \delta(\sigma)=\frac{1}{3AR^2(1+C_0R^2)^2}\left[\frac{\sqrt{A(1+C_0R^2)^3}-1}{\sqrt{A(1+C_0R^2)^3}+1}\right]^n\,f^{*}(C_0),
\end{equation}
where $n=0$ for the DMPR exterior solution and $n=1$ for the CFM exterior solution.

\begin{figure}
  \includegraphics[scale=0.7]{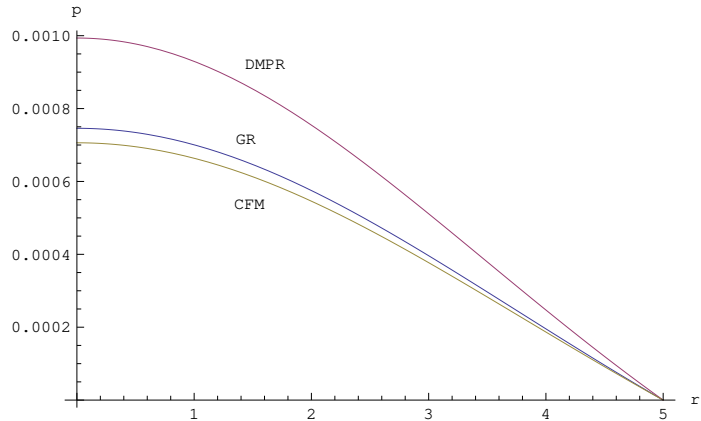}
\centering\caption{Qualitative comparison of the pressure for the same compact distribution ($A=0.3$) in general relativity (middle curve) and in the braneworld model with two different exteriors solutions: DMPR (upper curve) and CFM (lower curve). }
%\label{Press}       % Give a unique label
\end{figure}

\section{Conclusions}

In the context of the Randall-Sundrum braneworld, the consequences of bulk gravity on compact stellar distributions was studied through a new braneworld solution to Einstein's field equations. This solution was constructed from a well known spherically symmetric stellar solution for a perfect fluid in general relativity, namely, the Heintzmann solution. In order to generate the braneworld version containing the anisotropic effects necessary for realistic stellar models, the minimal geometric deformation approach was used to break the perfect fluid solution represented by the Heintzmann solution. 
Hence some important features of the five-dimensional consequences on physical variables inside compact distributions was clarified. In order to elucidate the specific role played by each exterior Weyl function, namely the dark radiation ${\cal U}^+$ and dark pressure ${\cal{ P}}^+$ on compact stellar systems,  the braneworld solution generated within the compact distribution was matched with two different exterior 
solutions, namely, the Dadhich, Maartens, Papadopoulos and Rezania solution (DMPR) and the Casadio, Fabbri and Mazzacurati solution (CFM). Both exterior solutions showed different effects on the stellar system. It was found that the DMPR solution produces an increase in the pressure and that the CFM reduces it. The fact that the CFM solution has no dark radiation, which is the main difference with the DMPR solution, 
seems to be the most likely cause of the opposite effects of these two exterior solutions on the stellar distribution. 

The physically acceptable solution developed in this paper represents the point of view of a
brane observer, hence it is not known whether the bulk eventually
constructed will be free of singularities. Despite the above, the important thing here is the fact that it was found a strong evidence showing that 
the exterior dark radiation  ${\cal U}^+$ always increases both the pressure and the compactness of the stellar structures,  and that the exterior dark pressure
${\cal P}^+$ always reduces them. If this is a general feature, it would mean that both exterior Weyl functions ${\cal U}^+$ and ${\cal P}^+$ have well 
defined consequences on stellar structure. Therefore an exterior solution with ${\cal U}^+=0$ and ${\cal P}^+\neq\,0$  surrounding a stellar 
distribution might be seen as an environment whose physical effects on the stellar structure are such that it can be considered as a region 
with negative effective pressure. This is certainly an interesting case which deserves further investigation.

\section*{Acknowledgments}

JO thanks Roberto Casadio for valuable discussions and Roy Maartens for useful comments. AS has been supported by project Nª 1121103 of Fondecyt.

\end{document}